\documentclass[preprint,amsmath,amssymb,aps,superscriptaddress]{revtex4}
\usepackage{graphicx}
\newsavebox{\paperfigurecaptionbox}
\newsavebox{\paperfiguregraphicbox}
\newlength{\paperfigureavailableheight}
\newcommand{\paperfigure}[3]{%
  \begin{lrbox}{\paperfigurecaptionbox}%
    \begin{minipage}{\linewidth}%
      \caption{#2}\label{#3}%
    \end{minipage}%
  \end{lrbox}%
  \setlength{\paperfigureavailableheight}{%
    \dimexpr\textheight-\ht\paperfigurecaptionbox
      -\dp\paperfigurecaptionbox-12pt\relax}%
  \sbox{\paperfiguregraphicbox}{%
    \includegraphics[width=\linewidth,height=\paperfigureavailableheight,
      keepaspectratio]{#1}}%
  \usebox{\paperfiguregraphicbox}%
  \par\nointerlineskip
  \usebox{\paperfigurecaptionbox}%
}

\usepackage{amsmath,amssymb}
\usepackage{ascmac}
\usepackage{url}
\usepackage{layout}
\usepackage{color}
\usepackage{xcolor}
\usepackage{braket}
\usepackage{physics}
\usepackage{mathrsfs}
\usepackage{booktabs}
\usepackage{here}
\usepackage{listings}
\usepackage{dcolumn}
\usepackage{fontspec}
\usepackage[noOT1]{newtxmath}
\let\hat\relax
\DeclareMathAccent{\hat}{\mathalpha}{largesymbols}{"62}

\usepackage{bm}

\usepackage[hidelinks]{hyperref}

\def\lesssim{\ \raise.3ex\hbox{$<$}\kern-0.8em\lower.7ex\hbox{$\sim$}\ }

\newcommand{\pdif}[2]{\frac{\partial#1}{\partial#2}}

\newcommand{\ncrex}[1]{\hat{\bm{\Psi}}_{#1}^\dagger}
\newcommand{\nannx}[1]{\hat{\bm{\Psi}}_{#1}}

\newcommand{\annp}[2]{\hat{#1}_{#2}}
\newcommand{\crep}[2]{\hat{#1}_{#2}^\dagger}

\begin{document}
\title{Specific heat in the BCS--rashbon BEC crossover regime of an ultracold Fermi gas with a 3D spherical spin-orbit coupling}
\author{Yusuke Okajima}
\affiliation{Department of Physics, Keio University, 3-14-1 Hiyoshi Kohoku-ku, Yokohama 223-8522, Japan}
\author{Koki Takemoto}
\affiliation{Department of Physics, Keio University, 3-14-1 Hiyoshi Kohoku-ku, Yokohama 223-8522, Japan}
\author{Shumpei Iwasaki}
\affiliation{Institute for Solid State Physics, University of Tokyo, Kashiwa, Chiba 277-8581, Japan}
\author{Yoji Ohashi}
\affiliation{Department of Physics, Keio University, 3-14-1 Hiyoshi Kohoku-ku, Yokohama 223-8522, Japan}
\begin{abstract}
  We theoretically investigate strong-coupling properties of an ultracold Fermi gas with a tunable spherical (isotropic) spin-orbit interaction. In this spin-orbit coupled system, it is known that a two-body bound state, called rashbon, can be formed even in the coupling regime where a pairing interaction alone cannot form such a two-body bound molecule. In this paper, we evaluate the specific heat at constant volume $C_V$, including pairing fluctuations in the presence of SOC within the framework of the strong-coupling theory developed by Nozi\`eres and Schmitt-Rink, to show that this thermodynamic quantity is very useful to see how system properties vary from a Fermi gas to a rashbon gas with increasing spin-orbit coupling strength. At the superfluid phase transition temperature $T_{\rm c}$, $C_V$ is found to exhibit a peak structure in the intermediate spin-orbit coupling regime between the weak-coupling Fermi gas regime and a strong-SOC rashbon gas regime. In the normal state above $T_{\rm c}$, from detailed analyses of the temperature dependences of $C_V$, we also identify the regions where system properties are dominated by (1) a Fermi atomic gas, (2) a Bose gas of stable rashbons, and (3) fluctuating rashbons. Combining these results with the conventional Bardeen-Cooper-Schrieffer (BCS)--Bose-Einstein condensation (BEC) crossover physics realized by tuning the strength of a pairing interaction, we present a unified description of the BCS--BEC and BCS--rashbon BEC crossover phenomena.
\end{abstract}
\maketitle
\par
\section{Introduction}
\par

High tunability is a major advantage of ultracold atomic gases. One example is a tunable pairing interaction associated with a Feshbach resonance~\cite{Chin_FeshbachResonance}. Using this interaction, the BCS (Bardeen-Cooper-Schrieffer)--BEC (Bose-Einstein condensation) crossover phenomenon~\cite{Bloch_UltracoldManyBodyReview,Chin_FeshbachResonance,Giorgini_FermiGasTheory,Randeria_Crossover,Ohashi_Crossover,Eagles_ElectronPairBinding,Leggett_BCSBECTheory} has been realized in $^{40}\mathrm{K}$~\cite{Regal_FermionPairCondensation} and $^6\mathrm{Li}$~\cite{Bartenstein_MolecularBEC,Zwierlein_MolecularBEC,Bourdel_CrossoverExperiment} Fermi gases, where a weak-coupling BCS superfluid of spatially extended Cooper pairs is continuously connected to a BEC of tightly bound molecules with increasing interaction strength~\cite{Nozieres_NSRTheory,SaDeMelo_FiniteTemperatureFluctuations,Ohashi_ResonantNSRTheory,Perali_PseudogapDiagnostics,Chen_PseudogapTheory,Hu_GaussianPairFluctuations,Haussmann_SelfConsistentTMatrix,Tsuchiya_HarmonicTrap}.

Another example is a synthetic tunable spin-orbit coupling (SOC)~\cite{Dalibard_SyntheticGaugeFields}: a 1D Raman-induced SOC was first realized in a Bose gas~\cite{Lin_1DSOCExperiment} and was subsequently implemented in degenerate $^{40}\mathrm{K}$~\cite{Wang_Fermionic1DSOC,Williams_ResonantSOC} and $^6\mathrm{Li}$~\cite{Cheuk_SpinInjectionSpectroscopy} Fermi gases. Recently, 2D synthetic SOCs have also been realized in $^{40}\mathrm{K}$~\cite{Huang_2DSOCExperiment,Meng_TopologicalBandGap} and $^{87}\mathrm{Sr}$~\cite{Liang_OpticalLattice2DSOC,Zhang_BeyondRotatingWaveSOC} Fermi gases. To the best of our knowledge, three-dimensional SOC has not yet been experimentally realized in an ultracold Fermi gas. Several theoretical schemes have been proposed for realizing 3D SOC~\cite{Anderson_3DSOCProposal,Anderson_MagneticGradientSOC}.

Since SOC is a one-body effect, it alone does not cause a superfluid instability in a Fermi gas. However, when 2D or 3D SOC is introduced into an attractively interacting Fermi gas, it can markedly affect pairing phenomena~\cite{Vyasanakere_TwoBodyBinding,Vyasanakere_Rashbons,Hu_AnisotropicSuperfluidity,Yu_ResonantSOCFermions,Gong_TopologicalTransition,Jiang_PairFluctuationFormalism,Iskin_AnisotropicSOCPhases,Chen_2DSOCCrossover,He_2DPairingEnhancement,He_SphericalSOCMass,Vyasanakere_CollectiveExcitations,He_FiniteTemperaturePairFluctuations,Yamaguchi_SphericalSOCNSR,Yamaguchi_ParityMixing}. In the two-body limit, a two-body bound state can form even in the weak pairing-interaction regime, where the pairing interaction alone cannot produce a bound state~\cite{Vyasanakere_TwoBodyBinding}. Such an SOC-assisted molecule is sometimes referred to as a rashbon in the literature.

In addition, starting from a Fermi superfluid in the weak-coupling BCS regime, increasing the strength of 2D or 3D SOC while keeping the weak pairing interaction fixed induces a phenomenon similar to the BCS--BEC crossover~\cite{Vyasanakere_Rashbons,Hu_AnisotropicSuperfluidity,Yu_ResonantSOCFermions,Gong_TopologicalTransition,Jiang_PairFluctuationFormalism,Iskin_AnisotropicSOCPhases,Chen_2DSOCCrossover,He_2DPairingEnhancement,He_SphericalSOCMass,Vyasanakere_CollectiveExcitations,He_FiniteTemperaturePairFluctuations,Yamaguchi_SphericalSOCNSR,Yamaguchi_ParityMixing}. In the strong-SOC regime, which corresponds to the strong-coupling BEC regime in the ordinary BCS--BEC crossover, the superfluid is well described by a BEC of rashbon molecules. This phenomenon is therefore sometimes referred to as the BCS--rashbon BEC crossover.

We note that the BCS--rashbon BEC crossover implies that, even when the pairing mechanism gives a very low superfluid transition temperature $T_{\rm c}$, $T_{\rm c}$ can be raised to $O(0.1T_{\rm F})$ by imposing a tunable 2D or 3D SOC on the system, where $T_{\rm F}$ is the Fermi temperature. At the current stage of cold Fermi gas physics, because the Feshbach-induced pairing mechanism is the only established pairing mechanism and because it is usually difficult to lower the temperature experimentally far below $T_{\rm F}$, this crossover phenomenon might be useful for exploring novel pairing mechanisms in ultracold Fermi gases beyond the Feshbach resonance.

In the ordinary BCS--BEC crossover realized by a tunable pairing interaction, not only superfluid properties but also normal-state properties continuously change from the normal Fermi gas regime to the normal Bose gas regime, through the pseudogap regime~\cite{vanWyk_SpecificHeatTemperatureDependence}. Thus, it is of interest to clarify how these three regions appear in the case of the BCS--rashbon BEC crossover, where the SOC strength is tuned while keeping the strength of a pairing interaction fixed. In particular, considering this problem in the phase diagram in terms of the SOC strength, pairing interaction strength, and temperature, we would be able to clarify how the BCS--rashbon BEC crossover is related to the ordinary BCS--BEC crossover.

In considering the normal-state properties of the BCS--rashbon BEC crossover, we briefly point out that the specific heat at constant volume $C_V$ is useful for characterizing whether the system is predominantly fermionic or bosonic. We note that $C_V$ increases with temperature in a Fermi gas but decreases with temperature in a Bose gas. Indeed, in the case of the ordinary BCS--BEC crossover, Ref.~\cite{vanWyk_SpecificHeatTemperatureDependence} identified the above-mentioned three regions by analyzing the detailed temperature dependence of this thermodynamic quantity.

In this paper, we theoretically investigate a two-component Fermi gas with both a tunable $s$-wave pairing interaction and a tunable 3D SOC. For simplicity, we deal with a spherical (isotropic) 3D SOC. To include effects of pairing fluctuations, we employ the strong-coupling theory developed by Nozi\`eres and Schmitt-Rink (NSR). Using this strong-coupling scheme, we evaluate $C_V$ at and above $T_{\rm c}$ in a wide parameter region with respect to the SOC strength and the pairing interaction strength. From detailed analyses of the temperature dependence of $C_V$, we clarify how normal-state properties vary in the BCS--rashbon BEC crossover phenomenon. We also identify the above-mentioned three regions in the phase diagram in terms of the SOC strength, pairing-interaction strength, and temperature, to present a unified description of the BCS--BEC and BCS--rashbon BEC crossovers.

This paper is organized as follows: In Sec.~II, we present a model of a spin-orbit-coupled interacting Fermi gas and explain our formalism within the framework of the NSR theory. In Sec.~III, we show the calculated $C_V$ at $T_{\rm c}$ and discuss the similarities and differences between the BCS--BEC and BCS--rashbon BEC crossovers. In Sec.~IV, we examine the normal-state properties above $T_{\rm c}$ in the BCS--rashbon BEC crossover regime through the temperature dependence of $C_V$. For simplicity, we set $\hbar=k_{\rm B}=1$ and take the system volume $V$ to be unity throughout this paper.

\par
\par
\section{Formulation}
\par
We consider a uniform two-component Fermi gas with a contact-type pairing interaction, as well as a spherical 3D SOC, described by the model Hamiltonian
\begin{align}
  H & =\sum_{\bm{p},\alpha}\xi_{\bm{p}}\crep{c}{\bm{p},\alpha} \annp{c}{\bm{p},\alpha} + \sum_{\bm{p},\alpha,\beta}\bm{p}_\lambda\cdot\bm{\sigma}_{\alpha\beta}\crep{c}{\bm{p},\alpha} \annp{c}{\bm{p},\beta}-U\sum_{\bm{p},\bm{p'},\bm{q}} \crep{c}{\bm{p}+\frac{\bm{q}}{2},\uparrow} \crep{c}{-\bm{p}+\frac{\bm{q}}{2},\downarrow} \annp{c}{-\bm{p'}+\frac{\bm{q}}{2},\downarrow} \annp{c}{\bm{p'}+\frac{\bm{q}}{2},\uparrow}.
  \label{hamiltonian}
\end{align}
Here, $\crep{c}{\bm p,\alpha}$ is the creation operator of a Fermi atom with momentum $\bm p$ and pseudospin $\alpha=\uparrow,\downarrow$, which labels two atomic hyperfine states. The kinetic energy $\xi_{\bm p}=\varepsilon_{\bm p}-\mu$ of a Fermi atom is measured from the Fermi chemical potential $\mu$, where $\varepsilon_{\bm p}=\bm p^2/(2m)$ and $m$ is the atomic mass. The quantity $-U~(<0)$ is the $s$-wave pairing interaction, which is assumed to be tunable. As usual, we measure the interaction strength in terms of the inverse scattering length $(k_{\rm F}a_{\rm s})^{-1}$ normalized by the Fermi momentum $k_{\rm F}$. The scattering length $a_{\rm s}$ is related to $U$ by
\begin{align}
  \frac{4\pi a_{\rm s}}{m} & =-\frac{U}{1-U\sum_{\bm{p}}^{p_{\rm c}}\frac{1}{2\varepsilon_{\bm{p}}}}.\label{renormalization}
\end{align}
Here, $p_{\rm c}$ is a cutoff momentum.

The second term in Eq.~(\ref{hamiltonian}) describes the spherical SOC, where $\bm\sigma=(\sigma_1,\sigma_2,\sigma_3)$ are Pauli matrices acting on pseudospin space, and
\begin{align}
  \bm{p}_\lambda=
  \begin{pmatrix}
    \lambda_x p_x \\   \lambda_y p_y   \\   \lambda_z p_z
  \end{pmatrix}.\label{411a-plamb}
\end{align}
Here, $\lambda_j~(j=x,y,z)$ are the spin-orbit coupling parameters. Since we deal with the spherical SOC, we set $\lambda_x=\lambda_y=\lambda_z=\lambda$ \cite{Vyasanakere_TwoBodyBinding,Vyasanakere_Rashbons,He_SphericalSOCMass}.

In the four-component Nambu representation, the model Hamiltonian in Eq.~(\ref{hamiltonian}) has the form
\begin{align}
  H & =\frac{1}{2}\sum_{\bm{p}} \ncrex{\bm{p}}\left[\xi_{\bm{p}}+\bm{p}_\lambda\cdot\bm{\alpha}\right](\tau_3\otimes\sigma_0) \nannx{\bm{p}} -\frac{U}{4}\sum_{\bm{q}}\rho_+\left(\frac{\bm{q}}{2}\right)\rho_-\left(-\frac{\bm{q}}{2}\right),\label{425-hamiltonian-nambu}
\end{align}
where
\begin{align}
  \nannx{\bm{p}} & =
  \begin{pmatrix}
    \annp{c}{\bm{p},\uparrow} \\   \annp{c}{\bm{p},\downarrow}   \\   \crep{c}{-\bm{p},\uparrow}  \\   \crep{c}{-\bm{p},\downarrow}
  \end{pmatrix}\label{nambu-field}
\end{align}
is the four-component Nambu field \cite{Nambu_Formalism}, and the components of
\begin{align}
  \bm{\alpha} & =\left(\tau_3\otimes\sigma_1,\,\tau_0\otimes\sigma_2,\,\tau_3\otimes\sigma_3\right)\label{nambu-spin-matrices}
\end{align}
are spin matrices in the Nambu representation. Here, $\bm\tau=(\tau_1,\tau_2,\tau_3)$ are the usual $2\times2$ Pauli matrices acting on particle-hole space, and $\tau_0$ and $\sigma_0$ are the identity matrices in particle-hole and pseudospin spaces, respectively. For example,
\begin{align}
  \alpha_1
  &=\tau_3\otimes\sigma_1
  =\begin{pmatrix}
    \sigma_1&0\\
    0&-\sigma_1
  \end{pmatrix}
  =\begin{pmatrix}
    0&1&0&0\\
    1&0&0&0\\
    0&0&0&-1\\
    0&0&-1&0
  \end{pmatrix}.\label{nambu-alpha-example}
\end{align}
In Eq.~(\ref{425-hamiltonian-nambu}),
\begin{align}
  \rho_\pm\left(\frac{\bm q}{2}\right)
  =\sum_{\bm p}
  \ncrex{\bm p+\frac{\bm q}{2}}
  \eta_\pm
  \nannx{\bm p-\frac{\bm q}{2}}
  \label{generalized-density}
\end{align}
with $\eta_\pm=[(\pm i\tau_1-\tau_2)\otimes\sigma_2]/2$, is a generalized density operator describing fluctuations in the Cooper channel. The single-particle dispersion measured from the chemical potential is
\begin{align}
  \xi_{\bm p}^{s}
  =\xi_{\bm p}+s|\bm p_\lambda|
  \equiv\varepsilon_{\bm p}^{s}-\mu,
  \qquad s=\pm1.
  \label{helicity-dispersion}
\end{align}
These single-particle states labeled by $s=\pm1$ are also referred to as the helicity states \cite{Vyasanakere_TwoBodyBinding,He_SphericalSOCMass}. We then find from the first term in Eq.~(\ref{425-hamiltonian-nambu}) that the $4\times4$ matrix single-particle thermal Green function in the absence of the pairing interaction ($U=0$) is given by
\begin{align}
  G_0(\bm p,i\omega_n)
  &=
  \frac{1}{i\omega_n-[\xi_{\bm p}+\bm p_\lambda\cdot\bm\alpha](\tau_3\otimes\sigma_0)}
  \nonumber\\
  &=-\frac{1}{2}\sum_{s=\pm1}
  \frac{1}{\omega_n^2+(\xi_{\bm p}^{s})^2}
  \left[i\omega_n+[\xi_{\bm p}+\bm p_\lambda\cdot\bm\alpha](\tau_3\otimes\sigma_0)\right]
  \left[\tau_0\otimes\sigma_0+s\frac{\bm p_\lambda\cdot\bm\alpha}{|\bm p_\lambda|}\right].
  \label{free-green-function}
\end{align}
Here, $\omega_n$ is the fermion Matsubara frequency.

We perturbatively include effects of the pairing interaction $-U$ within the framework of the NSR theory \cite{Nozieres_NSRTheory,Ohashi_ResonantNSRTheory,Yamaguchi_SphericalSOCNSR}. The thermodynamic potential in this strong-coupling scheme has the form
\begin{align}
  \Omega=\Omega_0+\Omega_{\rm NSR},
  \label{omega-decomposition}
\end{align}
where
\begin{align}
  \Omega_0
  =-T\sum_{\bm p,s=\pm1}
  \ln\left[1+e^{-\beta\xi_{\bm p}^{s}}\right],
  \label{omega-zero}
\end{align}
is the thermodynamic potential of a spin-orbit-coupled free Fermi gas with the single-particle dispersion $\xi_{\bm p}^{s}$ defined in Eq.~(\ref{helicity-dispersion}), where $\beta=1/T$.

The NSR correction term $\Omega_{\rm NSR}$ is diagrammatically given in Fig.~\ref{fig:nsr-ring-sum}. Summing over those diagrams, one has \cite{Nozieres_NSRTheory,vanWyk_SpecificHeatTemperatureDependence}
\begin{align}
  \Omega_{\rm NSR}
  =-T\sum_{\bm q,i\nu_n}
  e^{i\delta\nu_n}\ln\Gamma(\bm q,i\nu_n).
  \label{omega-nsr}
\end{align}
\begingroup
\begin{figure}[tbp]
  \centering
  \paperfigure{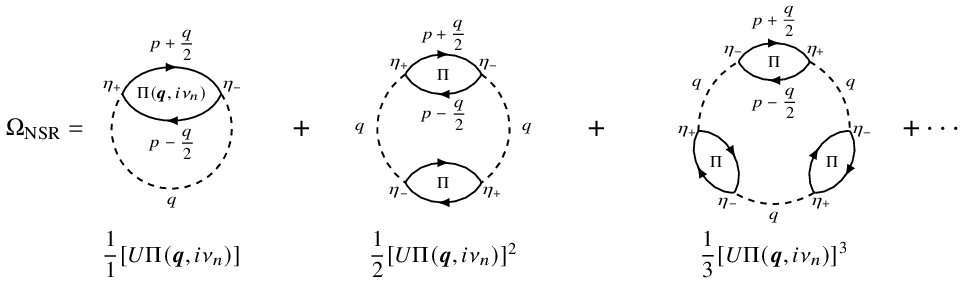}{NSR correction $\Omega_{\rm NSR}$ to the thermodynamic potential $\Omega=\Omega_0+\Omega_{\rm NSR}$. The solid and the dashed lines denote the single-particle thermal Green's function $G_0(\bm p,i\omega_n)$ in Eq.~(\ref{free-green-function}) and the pairing interaction $-U$, respectively. $\Pi(\bm q,i\nu_n)$ is the pair-correlation function in Eq.~(\ref{pair-correlation}).}{fig:nsr-ring-sum}
\end{figure}
\endgroup
Here, $\delta$ is an infinitesimally small positive number, and $\nu_n$ is the boson Matsubara frequency. In Eq.~(\ref{omega-nsr}),
\begin{align}
  \Gamma\left(\bm{q},i\nu_n\right) & =\frac{1}{\frac{m}{4\pi a_{\rm s}}+\left[\Pi\left(\bm{q},i\nu_n\right)-\sum_{\bm{p}}\frac{1}{2\varepsilon_{\bm{p}}}\right]}\label{tmatrix}
\end{align}
is the particle-particle scattering matrix describing pairing fluctuations, where
\begin{align}
   & \Pi\left(\bm{q},i\nu_n\right) = -\frac{1}{4}\sum_{\bm{p},s,s'}\frac{1-n_{\rm F}\left(\xi^{s}_{\bm{p}+\frac{\bm{q}}{2}}\right)-n_{\rm F}\left(\xi^{s'}_{\bm{p}-\frac{\bm{q}}{2}}\right)}{i\nu_n-\xi^{s}_{\bm{p}+\frac{\bm{q}}{2}}-\xi^{s'}_{\bm{p}-\frac{\bm{q}}{2}}}\nonumber \\
   & \hspace{40truemm}\times\left[1+ss'\frac{\left(\bm{p}+\frac{\bm{q}}{2}\right)_ \lambda \cdot \left(\bm{p}-\frac{\bm{q}}{2}\right)_\lambda}{\left|\left(\bm{p}+\frac{\bm{q}}{2}\right)_ \lambda\right|\left|\left(\bm{p}-\frac{\bm{q}}{2}\right)_\lambda\right|}\right]\label{pair-correlation}
\end{align}
is the pair-correlation function, with $n_{\rm F}(x)$ being the Fermi distribution function. In Eq.~(\ref{pair-correlation}), the factor in square brackets physically describes the projection of the two helicity states ($s,s'=\pm1$) onto the spin-singlet channel \cite{Vyasanakere_TwoBodyBinding,Vyasanakere_Rashbons,Hu_AnisotropicSuperfluidity,Yu_ResonantSOCFermions,Gong_TopologicalTransition,Jiang_PairFluctuationFormalism,Iskin_AnisotropicSOCPhases,Chen_2DSOCCrossover,He_2DPairingEnhancement,He_SphericalSOCMass,Yamaguchi_SphericalSOCNSR}.

As usual, the superfluid transition temperature $T_{\rm c}$ is determined from the Thouless criterion, stating that the superfluid instability occurs when the particle-particle scattering matrix $\Gamma(\bm q,i\nu_n)$ in Eq.~(\ref{tmatrix}) has a pole at $\bm q=\bm 0$ and $i\nu_n=0$ \cite{Nozieres_NSRTheory,Ohashi_ResonantNSRTheory,Haussmann_SelfConsistentTMatrix,Yamaguchi_SphericalSOCNSR}. The resulting $T_{\rm c}$ equation is given by \cite{Yamaguchi_SphericalSOCNSR}
\begin{align}
  0
  =\frac{m}{4\pi a_{\rm s}}+
  \sum_{\bm p}
  \left[
    \frac{\tanh\!\left(\xi_{\bm p}^{+}/2T\right)}{4\xi_{\bm p}^{+}}
    +\frac{\tanh\!\left(\xi_{\bm p}^{-}/2T\right)}{4\xi_{\bm p}^{-}}
    -\frac{1}{2\varepsilon_{\bm p}}
  \right].
  \label{thouless}
\end{align}
At $T_{\rm c}$, we solve this $T_{\rm c}$ equation, together with the equation for the number of Fermi atoms,
\begin{align}
  N
  =-\left(\frac{\partial\Omega}{\partial\mu}\right)_{T}
  =N_{\rm F0}+N_{\rm NSR},
  \label{number-eq}
\end{align}
where
\begin{align}
  N_{\rm F0}
  &=-\left(\frac{\partial\Omega_0}{\partial\mu}\right)_{T}
  =\sum_{\bm p,s}n_{\rm F}\!\left(\xi_{\bm p}^{s}\right),
  \label{number-free-fermi}
\end{align}
is the number of non-interacting (but spin-orbit-coupled) Fermi atoms, and
\begin{align}
  N_{\rm NSR}
  &=-T\sum_{\bm q,i\nu_n}e^{i\delta\nu_n}
  \Gamma(\bm q,i\nu_n)
  \frac{\partial}{\partial\mu}\Pi(\bm q,i\nu_n)
  \label{number-nsr}
\end{align}
is the NSR correction, to determine $T_{\rm c}$ and $\mu$ self-consistently. Although their self-consistent solutions have been obtained in Ref.~\cite{Yamaguchi_SphericalSOCNSR}, we recalculate them here and briefly show the results in Fig.~\ref{tc-mu-self-consistent} for readers' convenience. These solutions are used to evaluate $C_V(T_{\rm c})$ in Sec.~III. Above $T_{\rm c}$, we only deal with the number equation~(\ref{number-eq}), to obtain $\mu(T>T_{\rm c})$.

Once $T_{\rm c}$ and $\mu(T\geq T_{\rm c})$ are obtained, we then evaluate the specific heat at constant volume $C_V$, given by
\begin{align}
  C_V=\left(\frac{\partial E}{\partial T}\right)_{V,N}.
  \label{specific-heat}
\end{align}
Here, the internal energy $E$ is obtained from the NSR thermodynamic potential $\Omega$ in Eq.~(\ref{omega-decomposition}) via the Legendre transformation as \cite{vanWyk_SpecificHeatTemperatureDependence},
\begin{align}
  E
  ={}&\Omega
  -T\left(\frac{\partial\Omega}{\partial T}\right)_{\mu}
  -\mu\left(\frac{\partial\Omega}{\partial\mu}\right)_{T}
  \nonumber\\
  ={}&\sum_{\bm p,s}\varepsilon_{\bm p}^{s}
  n_{\rm F}\!\left(\xi_{\bm p}^{s}\right)
  \nonumber\\
  &-T\sum_{\bm q,i\nu_n}e^{i\delta\nu_n}
  \Gamma(\bm q,i\nu_n)
  \left[
    T\frac{\partial}{\partial T}\Pi(\bm q,i\nu_n)
    +\mu\frac{\partial}{\partial\mu}\Pi(\bm q,i\nu_n)
  \right].
  \label{energy}
\end{align}
In this paper, we numerically carry out the $T$ derivative in Eq.~(\ref{specific-heat}).
\begin{figure}[tbp]
  \centering
  \paperfigure{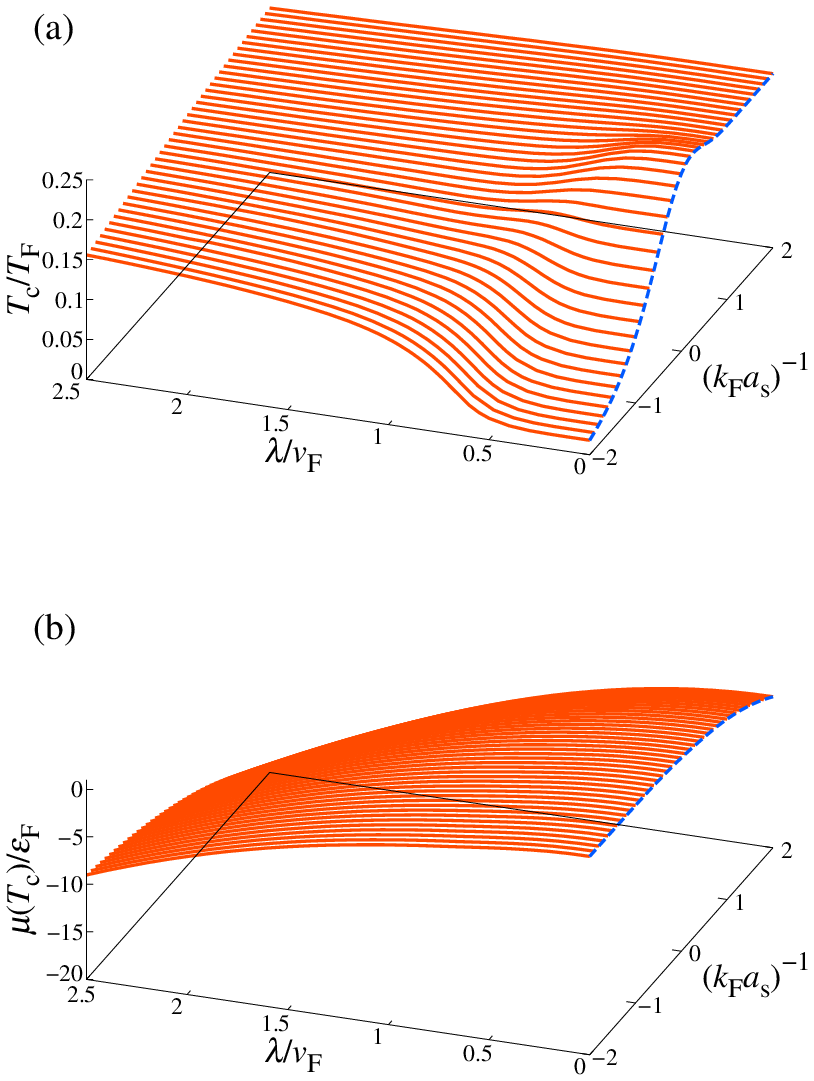}{Self-consistent solutions of the $T_{\rm c}$ equation~(\ref{thouless}) coupled with the number equation~(\ref{number-eq}). (a) $T_{\rm c}$, (b) $\mu(T_{\rm c})$. The dashed curves show the solutions at $\lambda=0$ in both panels. $T_{\rm F}$, $\varepsilon_{\rm F}$, and $v_{\rm F}$ are the Fermi temperature, Fermi energy, and Fermi velocity, respectively.}{tc-mu-self-consistent}
\end{figure}

Nozi\`eres and Schmitt-Rink pointed out that the NSR correction term $N_{\rm NSR}$ may be decomposed as~\cite{Nozieres_NSRTheory}
\begin{align}
  N_{\rm NSR}=2N_{\rm st}+N_{\rm sc}.
  \label{nst-nsc-decomp}
\end{align}
Here,
\begin{align}
  N_{\rm st}
  =\frac{1}{2}\sum_{\bm q:{\rm pole}}
  n_{\rm B}(\omega_{\bm q})
  \left.
  \frac{
    \dfrac{\partial}{\partial\mu}
    \Pi(\bm q,\omega+i\delta)
  }{
    \dfrac{\partial}{\partial\omega}
    \Pi(\bm q,\omega+i\delta)
  }
  \right|_{\omega=\omega_{\bm q}}
  \label{nst-pole-number}
\end{align}
is the number of stable pairs, where $n_{\rm B}(x)=[\exp(x/T)-1]^{-1}$ is the Bose distribution function and $\omega$ is a real frequency. The $\bm q$ summation is taken over real poles $\omega_{\bm q}$ of the analytically continued particle-particle scattering matrix $\Gamma(\bm q,\omega+i\delta)$ below the two-particle continuum threshold. The derivation is given in Appendix~\ref{app:nst-nsc}. The factor of two in Eq.~(\ref{nst-nsc-decomp}) reflects the fact that each stable pair consists of two fermionic atoms.

The remaining contribution, $N_{\rm sc}$, is the atom-number contribution of scattering states and includes metastable pairing fluctuations. It is evaluated as
\begin{align}
  N_{\rm sc}=N_{\rm NSR}-2N_{\rm st}.
  \label{nsc-number-remainder}
\end{align}
Here, $N_{\rm sc}$ is an interaction correction to the atom number, rather than a positive-definite population of metastable molecules, and can therefore be negative~\cite{Nozieres_NSRTheory}.
Although this decomposition was originally introduced in ordinary BCS--BEC crossover physics, it is also applicable to the BCS--rashbon BEC crossover. In this paper, $N_{\rm st}$ denotes the number of stable molecules at $\lambda=0$ and the number of stable rashbons for the extension of these pairs to $\lambda>0$.

Before concluding this section, we note that there is no clear boundary between the molecular bosons in the ordinary BCS--BEC crossover and the rashbons in the BCS--rashbon BEC crossover. Distinguishing stable molecules at $\lambda=0$ from stable rashbons at $\lambda>0$ is a convention adopted in this paper and does not imply a phase transition between them.
\par
\section{Specific heat $C_V$ in the BCS--rashbon BEC crossover region at $T_{\rm c}$}
\par
\begin{figure}[tbp]
  \centering
  \paperfigure{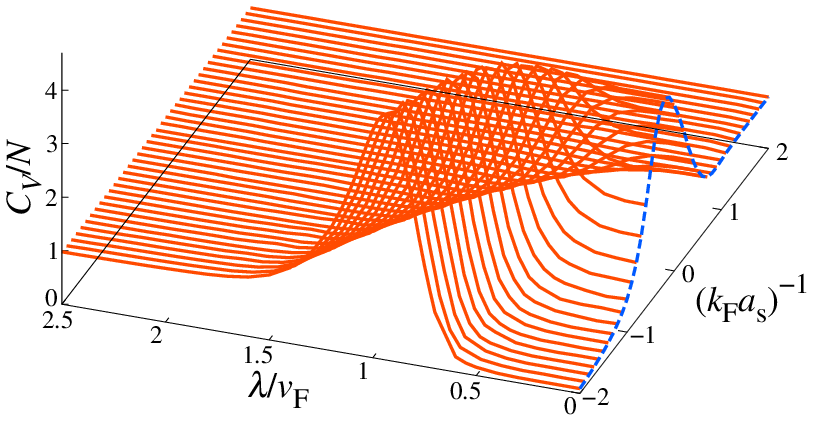}{Calculated specific heat per particle $C_V/N$ at $T_{\rm c}$, as a function of the $s$-wave interaction strength measured in terms of the inverse scattering length $(k_{\rm F}a_{\rm s})^{-1}$ and the spin-orbit coupling strength $\lambda/v_{\rm F}$. The dashed curve shows the result at $\lambda=0$. $k_{\rm F}$ and $v_{\rm F}$ are the Fermi momentum and Fermi velocity, respectively.}{sheat_lamas}
\end{figure}
\par
Figure~\ref{sheat_lamas} shows the NSR result for $C_V/N$ at $T_{\rm c}$. In the weak-coupling BCS regime when $(k_{\rm F}a_{\rm s})^{-1}=-2$, the dependence of $C_V(T_{\rm c})$ on the SOC strength $\lambda$ is found to qualitatively be similar to the BCS--BEC crossover behavior of this quantity seen at $\lambda=0$. In particular, the fact that the peak structures in these two cases smoothly connect in this figure suggests that the two peaks have a closely related origin.

\begin{figure}[tbp]
  \centering
  \paperfigure{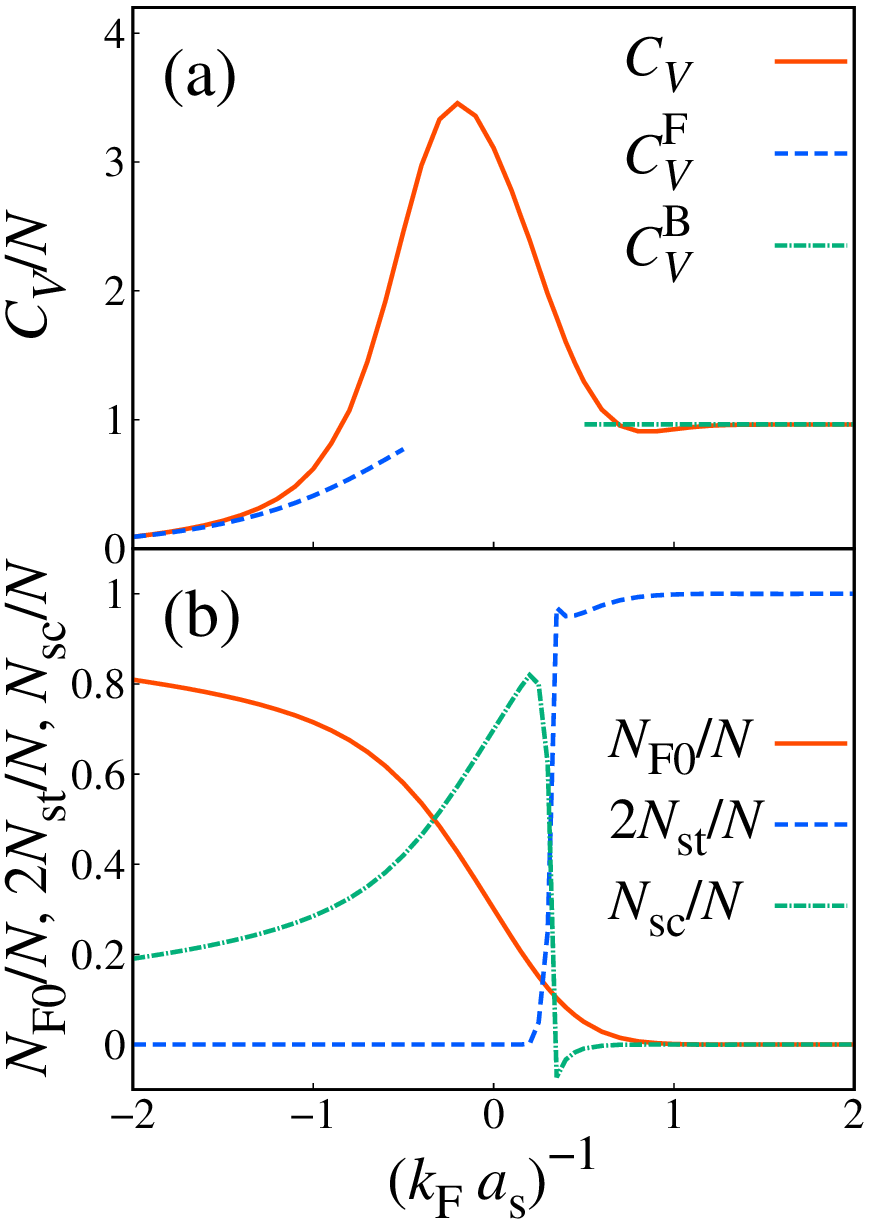}{(a) $C_V(T_{\rm c})$ as a function of the pairing-interaction strength $(k_{\rm F}a_{\rm s})^{-1}$ in the absence of SOC ($\lambda=0$). $C_V^{\rm F}(T_{\rm c})$ is the specific heat of a free Fermi gas at $T_{\rm c}$, given in Eq.~(\ref{cv-free-fermi}). $C_V^{\rm B}(T_{\rm BEC})$ is the specific heat of an ideal gas of $N/2$ bosons with a mass $M_{\rm B}=2m$ at the BEC phase-transition temperature $T_{\rm BEC}$, given in Eq.~(\ref{cv-ideal-bose}). (b) The numbers of unpaired Fermi atoms $N_{\rm F0}$ in Eq.~(\ref{number-free-fermi}), stable two-body bound molecules $N_{\rm st}$, and the contribution of scattering states $N_{\rm sc}$, defined in the discussion following Eq.~(\ref{nst-nsc-decomp}), in the ordinary BCS--BEC crossover regime at $T_{\rm c}$.}{tc-interaction-scan}
\end{figure}

Figure~\ref{tc-interaction-scan}(a) shows $C_V(T_{\rm c})$ extracted from Fig.~\ref{sheat_lamas} when $\lambda=0$. In the weak-coupling BCS regime ($(k_{\rm F}a_{\rm s})^{-1}\lesssim-1$), $C_V(T_{\rm c})$ is found to agree well with the specific heat $C_V^{\rm F}(T_{\rm c})$ of a free Fermi gas, given by
\begin{align}
  C_V^{\rm F}(T_{\rm c})=\frac{\pi^2}{2}\frac{T_{\rm c}}{T_{\rm F}}N.
  \label{cv-free-fermi}
\end{align}
Here, $T_{\rm F}$ is the Fermi temperature. Indeed, Fig.~\ref{tc-interaction-scan}(b) indicates that the dominant component is the contribution from unpaired Fermi atoms. As the pairing-interaction strength increases, Fig.~\ref{tc-interaction-scan}(a) shows that $C_V(T_{\rm c})$ develops a peak. In the strong-coupling BEC regime, $(k_{\rm F}a_{\rm s})^{-1}\gtrsim+1$, where most Fermi atoms form tightly bound molecules (see Fig.~\ref{tc-interaction-scan}(b)), $C_V(T_{\rm c})$ eventually approaches that of an ideal gas of molecules with molecular mass $M_{\rm B}=2m$:
\begin{align}
  C_V^{\rm B}(T_{\rm BEC})
  =\frac{15}{4}\frac{\zeta(5/2)}{\zeta(3/2)}\left(\frac{N}{2}\right)
  \simeq0.963N.
  \label{cv-ideal-bose}
\end{align}
Here, $\zeta(x)$ is the Riemann zeta function.
The scattering-state contribution $N_{\rm sc}$, which includes fluctuating pairs, is large in the region where $C_V(T_{\rm c})$ reaches its maximum (see Fig.~\ref{tc-interaction-scan}(b)), suggesting that the peak is related to the enhanced bosonic character of the system. We discuss this point later together with the corresponding behavior in the BCS--rashbon BEC crossover.

\begin{figure}[tbp]
  \centering
  \paperfigure{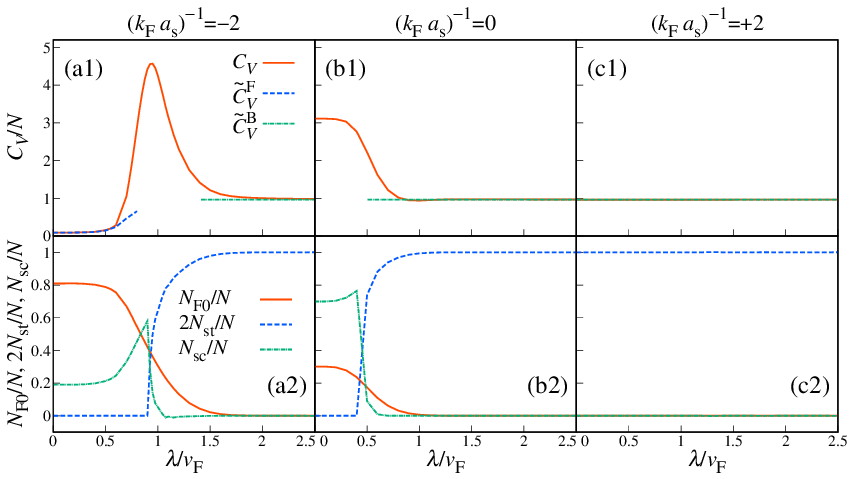}{The upper panels (a1)--(c1) show $C_V(T_{\rm c})$ as a function of the SOC strength $\lambda/v_{\rm F}$. In each case, $N_{\rm F0}$, $N_{\rm st}$, and $N_{\rm sc}$ are shown in the lower panels (a2)--(c2). Panels (a1, a2), (b1, b2), and (c1, c2) correspond to $(k_{\rm F}a_{\rm s})^{-1}=-2$, 0, and $+2$, respectively. $\widetilde{C}_V^{\rm F}(T_{\rm c})$ is the specific heat of a free Fermi gas in the presence of SOC ($\lambda\ne0$), and $\widetilde{C}_V^{\rm B}(\widetilde{T}_{\rm BEC})$ is that of an ideal gas of $N/2$ rashbons.}{tc-soc-scans}
\end{figure}

We proceed to the case of the BCS--rashbon BEC crossover. Figure~\ref{tc-soc-scans}(a1) shows $C_V(T_{\rm c})$ as a function of the SOC strength $\lambda$ when $(k_{\rm F}a_{\rm s})^{-1}=-2$, where the pairing interaction is weak. In the weak-SOC regime, $\lambda/v_{\rm F}\ll1$, the behavior of $C_V(T_{\rm c})$ is well described by the specific heat $\widetilde{C}_V^{\rm F}$ of a spin-orbit-coupled free Fermi gas, given by
\begin{align}
  \widetilde{C}_V^{\rm F}
  =
  \left[
    \frac{\partial}{\partial T}
    \sum_{\bm p,s=\pm}
    \varepsilon_{\bm p}^{s}
    n_{\rm F}\!\left(\xi_{\bm p}^{s}\right)
  \right]_{V,N}.
  \label{cv-soc-free-fermi}
\end{align}
Here, the Fermi chemical potential $\mu$ is determined from the number equation
\begin{align}
  N
  =
  \sum_{\bm p,s}
  n_{\rm F}\!\left(\xi_{\bm p}^{s}\right).
  \label{number_of_fermi}
\end{align}
Starting from the weak-SOC regime, Figs.~\ref{tc-soc-scans}(a1, a2) show that both $C_V(T_{\rm c})$ and the scattering-state contribution $N_{\rm sc}$, which includes metastable pairing fluctuations, increase with increasing SOC strength $\lambda$, although the pairing-interaction strength remains fixed at $(k_{\rm F}a_{\rm s})^{-1}=-2$ in the weak-coupling BCS regime.

Thus, as in the ordinary BCS--BEC crossover case shown in Fig.~\ref{tc-interaction-scan}, the enhancement of the bosonic character of the system is crucial for understanding the origin of the peak structure in Fig.~\ref{tc-soc-scans}(a1).

As the SOC strength is increased toward the strong-SOC limit ($\lambda/v_{\rm F}\gg1$), Fig.~\ref{tc-soc-scans}(a2) shows that $N_{\rm st}$ approaches $N/2$, indicating that most Fermi atoms form stable rashbons. Indeed, as shown in Appendix~\ref{app:strong-coupling}, the internal energy in Eq.~(\ref{energy}) and the number equation~(\ref{number-eq}) are, respectively, reduced to the expressions for an ideal rashbon Bose gas. Its specific heat is
\begin{align}
  \widetilde{C}_V^{\rm B}
  =
  \left(\pdif{E}{T}\right)_{V,N}.
  \label{cv-rashbon-bose}
\end{align}
The internal energy and number equation are
\begin{align}
  E
  &=
  \sum_{\bm q}
  \frac{\bm q^2}{2M_{\rm RB}}
  n_{\rm B}\!\left(
    \frac{\bm q^2}{2M_{\rm RB}}-\mu_{\rm RB}
  \right)
  -E_{\rm bind}^{\rm RB}\frac{N}{2},
  \label{energy_of_rb}
  \\
  \frac{N}{2}
  &=
  \sum_{\bm q}
  n_{\rm B}\!\left(
    \frac{\bm q^2}{2M_{\rm RB}}-\mu_{\rm RB}
  \right).
  \label{number_of_rb}
\end{align}
Here, the rashbon chemical potential, binding energy, and effective mass are given by~\cite{He_SphericalSOCMass}
\begin{align}
  \mu_{\rm RB}&=2\mu+E_{\rm bind}^{\rm RB},\label{rashbon-chemical-potential}\\
  E_{\rm bind}^{\rm RB}&=m\lambda^2J,\label{rashbon-binding-energy}\\
  M_{\rm RB}&=\frac{2m}{\dfrac{7}{3}-\dfrac{2}{J}
    -\dfrac{4}{3}\left[\dfrac{J-1}{J}\right]^{3/2}},\label{rashbon-effective-mass}\\
  J&=1+\frac14\left[\frac{1}{m\lambda a_{\rm s}}
    +\sqrt{\left[\frac{1}{m\lambda a_{\rm s}}\right]^2+4}\right]^2.
  \label{rashbon-j-definition}
\end{align}

Thus, as seen in Fig.~\ref{tc-soc-scans}(a1), $C_V(T_{\rm c})$ in this regime approaches the specific heat $\widetilde{C}_V^{\rm B}(\widetilde{T}_{\rm BEC})$ of this rashbon Bose gas at the BEC transition temperature $\widetilde{T}_{\rm BEC}$:
\begin{align}
  \widetilde{C}_V^{\rm B}(\widetilde{T}_{\rm BEC})
  &=
  \left.\left(
    \pdif{E}{T}
  \right)_{V,N}\right|_{\mu_{\rm RB}=0}
  \nonumber\\
  &=
  \int\frac{dq\,q^2}{2\pi^2}
  \left(
    \frac{\bm q^2}{2M_{\rm RB}\widetilde{T}_{\rm BEC}}
  \right)^2
  \frac{
    \exp\!\left(
      \dfrac{\bm q^2}{2M_{\rm RB}\widetilde{T}_{\rm BEC}}
    \right)
  }{
    \left[
      \exp\!\left(
        \dfrac{\bm q^2}{2M_{\rm RB}\widetilde{T}_{\rm BEC}}
      \right)-1
    \right]^2
  }
  \nonumber\\
  &\simeq0.963N.
  \label{cvrb}
\end{align}

Equation~(\ref{cvrb}) gives the same value as $C_V^{\rm B}(T_{\rm BEC})$ in Eq.~(\ref{cv-ideal-bose}), obtained in the strong-coupling BEC limit of the ordinary BCS--BEC crossover. Although the rashbon mass $M_{\rm RB}$ differs from the molecular mass $M_{\rm B}=2m$ in the latter BEC regime, the BEC transition temperature also changes from $T_{\rm BEC}$ to $\widetilde{T}_{\rm BEC}$. Since $\widetilde{T}_{\rm BEC}\propto M_{\rm RB}^{-1}$, their product in the integrand of Eq.~(\ref{cvrb}) is independent of $M_{\rm RB}$, and hence $\widetilde{C}_V^{\rm B}(\widetilde{T}_{\rm BEC})=C_V^{\rm B}(T_{\rm BEC})$. Indeed, $C_V(T_{\rm c})$ is almost $\lambda$ independent in the strong-SOC rashbon BEC regime, as shown in Fig.~\ref{tc-soc-scans}(a1), although the rashbon effective mass $M_{\rm RB}$ depends on $\lambda$. Therefore, $C_V(T_{\rm c})$ does not distinguish the BEC and rashbon BEC limits once the system is already dominated by stable pairs.

Figure~\ref{tc-soc-scans}(b1) shows the case in which SOC is increased from the unitary limit, $(k_{\rm F}a_{\rm s})^{-1}=0$. In this case, pairing fluctuations described by $N_{\rm sc}$ are soon replaced by stable rashbons, as seen in Fig.~\ref{tc-soc-scans}(b2), so that $C_V(T_{\rm c})$ similarly decreases toward $\widetilde{C}_V^{\rm B}(\widetilde{T}_{\rm BEC})$ with increasing $\lambda$.

In Figs.~\ref{tc-soc-scans}(c1, c2), where $(k_{\rm F}a_{\rm s})^{-1}=+2$, stable molecules already dominate at $\lambda=0$, and $C_V(T_{\rm c})$ remains close to the ideal-Bose-gas value $0.963N$ throughout the plotted SOC range.

To grasp the common background physics behind the peak structures in Figs.~\ref{tc-interaction-scan}(a) and \ref{tc-soc-scans}(a1), it is helpful to consider the thermal dissociation of ``molecules.'' Since the total number $N$ of Fermi atoms is conserved, differentiating Eq.~(\ref{number-eq}) with respect to temperature gives
\begin{align}
  0
  =
  \left(
    \pdif{N_{\rm F0}}{T}
  \right)_{V,N}
  +
  \left(
    \pdif{N_{\rm NSR}}{T}
  \right)_{V,N}.
  \label{number-temperature-derivative}
\end{align}
On the other hand, if $N_{\rm NSR}$ is simply regarded as the number of Fermi atoms forming stable and fluctuating ``molecules,'' and $N_{\rm NSR}/2$ as the number of such ``molecules,'' the effective number of particles
\begin{align}
  \widetilde{N}
  =
  N_{\rm F0}
  +
  \frac{N_{\rm NSR}}{2}
  \label{effective-particle-number}
\end{align}
is not conserved on heating, where a ``molecule'' is also counted as one particle. We therefore define
\begin{align}
  R(T)
  &\equiv
  \left(
    \pdif{\widetilde{N}}{T}
  \right)_{V,N}
  \nonumber\\
  &=
  \left(
    \pdif{N_{\rm F0}}{T}
  \right)_{V,N}
  +
  \frac{1}{2}
  \left(
    \pdif{N_{\rm NSR}}{T}
  \right)_{V,N}
  \nonumber\\
  &=
  \frac{1}{2}
  \left(
    \pdif{N_{\rm F0}}{T}
  \right)_{V,N},
  \label{particle-production-rate}
\end{align}
where Eq.~(\ref{number-temperature-derivative}) has been used in the last line. Noting that the specific heat at constant volume is related to the entropy $S$ as
\begin{align}
  C_V(T)
  =
  T
  \left(
    \pdif{S}{T}
  \right)_{V,N},
  \label{specific-heat-entropy}
\end{align}
we expect the ``particle production'' on heating described by $R(T)$ to affect $C_V(T)$ through the temperature dependence of the entropy. Indeed, Fig.~\ref{fig:tc-cv-n0-response} shows that the peak position of $R(T_{\rm c})$ lies very close to that of $C_V(T_{\rm c})$ in both the BCS--BEC and BCS--rashbon BEC crossover cases. Thus, in both crossover phenomena, the proximity of the peaks in $R(T_{\rm c})$ and $C_V(T_{\rm c})$ supports the interpretation that rapid thermal redistribution from the correlated-pair contribution $N_{\rm NSR}$ to the unpaired-atom contribution $N_{\rm F0}$ is responsible for the specific-heat enhancement.

\begin{figure}[tbp]
  \centering
  \paperfigure{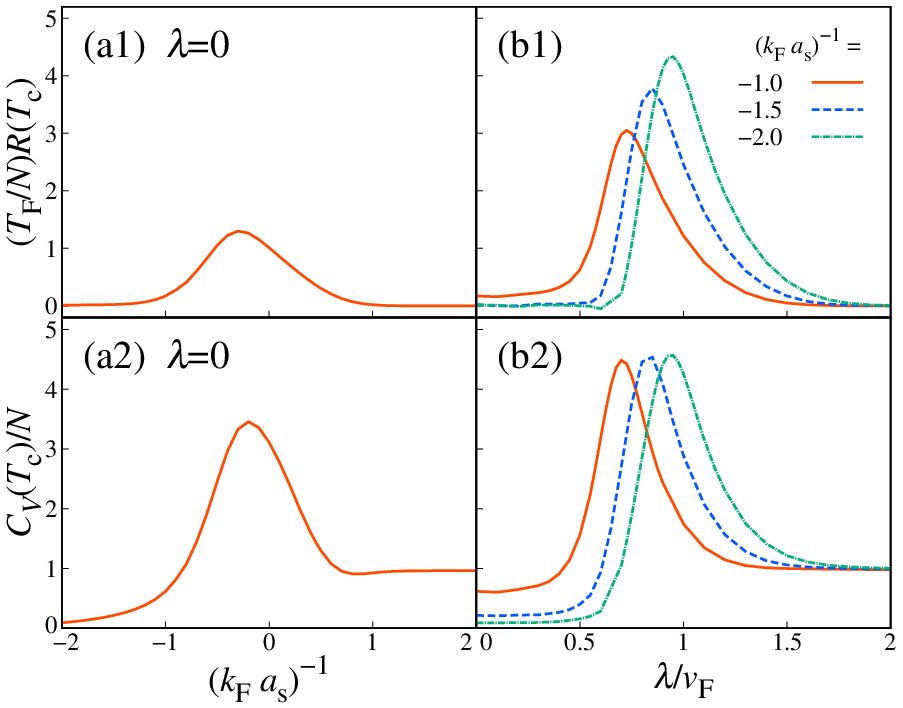}{Calculated $R(T_{\rm c})$ given in Eq.~(\ref{particle-production-rate}) in the BCS--BEC crossover case (a1) and the BCS--rashbon BEC crossover case (b1). In each case, $C_V(T_{\rm c})$ is also shown in the lower panel.}{fig:tc-cv-n0-response}
\end{figure}

To grasp more detailed background physics of the peak, we again plot $C_V(T_{\rm c})$ in Fig.~\ref{fig:tc-effective-mu-components}, but now as a function of the Fermi chemical potential $\widetilde{\mu}$, measured from the bottom of the band. That is, since the lowest energy of the helicity bands $\varepsilon_{\bm p}^{\pm}$ in Eq.~(\ref{helicity-dispersion}) equals $-m\lambda^2/2$ ($\leq0$), $\widetilde{\mu}$ is given by
\begin{align}
  \widetilde{\mu}
  =
  \begin{cases}
    \mu(T_{\rm c}) &: \text{BCS--BEC crossover},\\
    \mu(T_{\rm c})+\dfrac{m}{2}\lambda^2 &: \text{BCS--rashbon BEC crossover}.
  \end{cases}
  \label{mutilde-def}
\end{align}
We briefly note that $\mu(T_{\rm c})$ monotonically decreases as one moves from the weak-coupling BCS regime to both the strong-coupling BEC regime and rashbon BEC regime (see Fig.~\ref{tc-mu-self-consistent}(b)). For the parameter paths considered here, $\widetilde{\mu}$ also decreases monotonically. A decrease of $\widetilde{\mu}$ in Fig.~\ref{fig:tc-effective-mu-components} therefore corresponds to an increase of the pairing-interaction (SOC) strength in the BCS--BEC (BCS--rashbon BEC) crossover case.

Figure~\ref{fig:tc-effective-mu-components} shows that, in the BCS--BEC crossover case ($\lambda=0$), $C_V(T_{\rm c})$ has a peak at $\widetilde{\mu}$ [$=\mu(T_{\rm c})$] $\sim0.5\varepsilon_{\rm F}$ (where $\varepsilon_{\rm F}$ is the Fermi energy), where the ``molecular contribution'' described by $N_{\rm NSR}$ is actually dominated by fluctuating pairs described by the scattering contribution $N_{\rm sc}$. In the BCS--rashbon BEC crossover, on the other hand, the peaks lie near $\widetilde\mu=0$, close to the region in which the scattering-state contribution $N_{\rm sc}$ is replaced by the stable-rashbon contribution $2N_{\rm st}$ as $\widetilde\mu$ decreases. Thus, together with the previous discussion in Fig.~\ref{fig:tc-cv-n0-response}, one finds that, while the thermal dissociation of ``molecules'' on heating is commonly the key to produce the peak structure in $C_V(T_{\rm c})$ in both crossover phenomena, the detailed molecular characteristics are quite different in the two crossover cases.

\begin{figure}[tbp]
  \centering
  \paperfigure{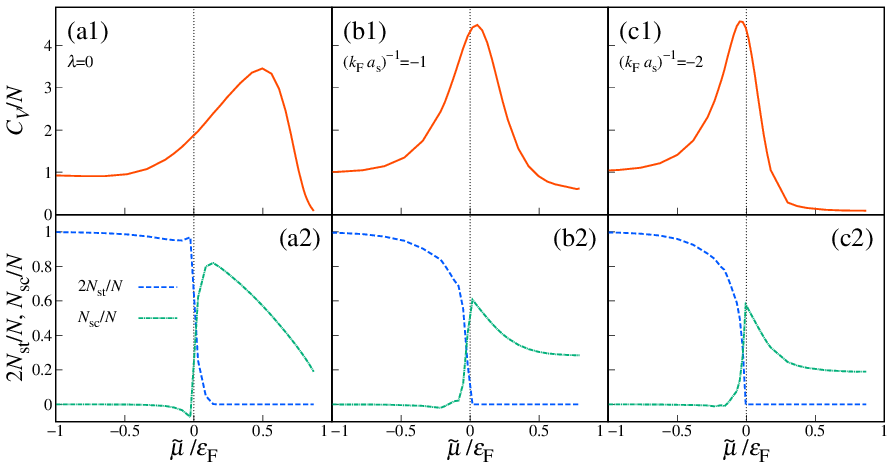}{Calculated $C_V$, $N_{\rm st}$, and $N_{\rm sc}$ at $T_{\rm c}$, as functions of $\widetilde{\mu}$ given in Eq.~(\ref{mutilde-def}). $C_V$ is shown in (a1)--(c1), and $N_{\rm st}$ and $N_{\rm sc}$ in (a2)--(c2). Panels (a1, a2) show the BCS--BEC crossover case ($\lambda=0$), while (b1, b2) and (c1, c2) show the BCS--rashbon BEC crossover cases at $(k_{\rm F}a_{\rm s})^{-1}=-1$ and $-2$, respectively. The vertical lines mark $\widetilde\mu=0$. $\varepsilon_{\rm F}$ is the Fermi energy.}{fig:tc-effective-mu-components}
\end{figure}

\begin{figure}[tbp]
  \centering
  \paperfigure{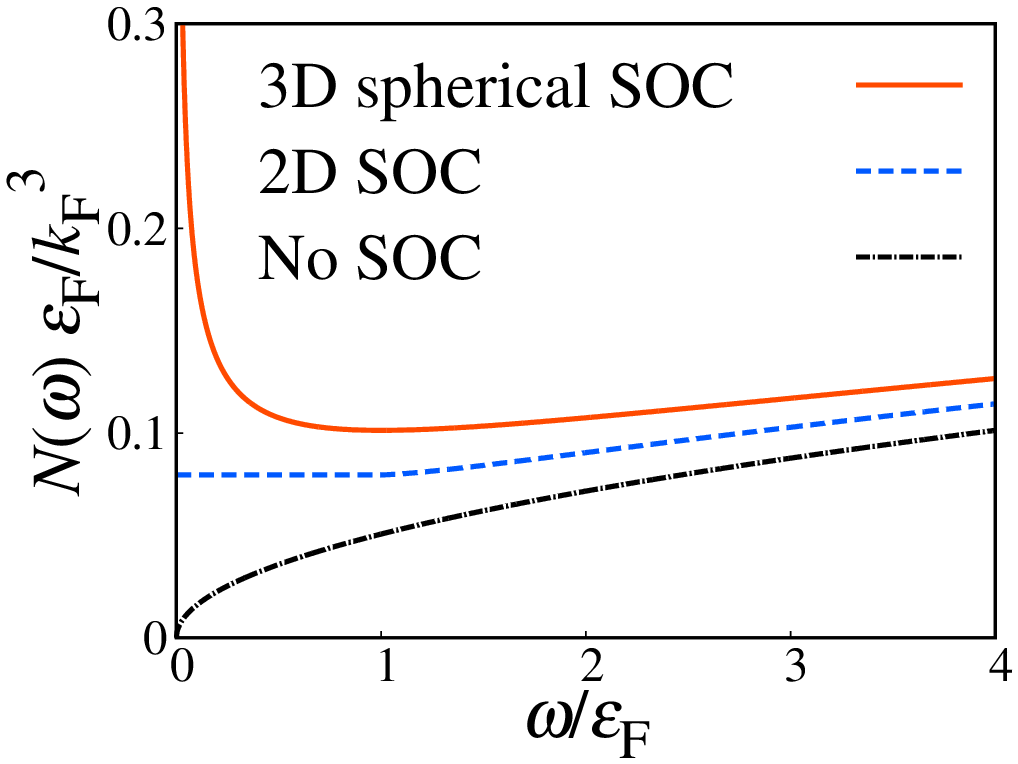}{``3D spherical SOC'' shows the single-particle density of states (DOS) $N(\omega)$ in a non-interacting Fermi gas with the 3D spherical SOC ($\lambda/v_{\rm F}=1$). The DOS is given per unit volume and summed over both helicity branches. In this figure, the energy $\omega$ is measured from the bottom of the energy band. ``2D SOC'' shows the DOS in a 3D non-interacting Fermi gas with the 2D SOC ($\lambda_x=\lambda_y=\lambda$, $\lambda_z=0$ in Eq.~(\ref{411a-plamb})). The 2D-SOC curve also has $\lambda/v_{\rm F}=1$. ``No SOC'' shows the DOS in a free Fermi gas ($\lambda=0$).}{fig:single-particle-dos}
\end{figure}

In the present spherical SOC case, the single-particle density of states (DOS) diverges as $N(\omega)\propto\omega^{-1/2}$ at the band bottom~\cite{He_SphericalSOCDOS}, as shown in Fig.~\ref{fig:single-particle-dos}. Thus, even when the pairing-interaction strength is weak, pairing fluctuations are enhanced by the large intensity of the DOS at $\omega=0$ when $\widetilde{\mu}=0$. In addition, as $\widetilde{\mu}$ decreases below zero, the $\bm q=0$ pair pole becomes separated from the two-particle continuum, and the stable-pole contribution $N_{\rm st}$ rapidly becomes dominant. These naturally explain the behavior of $N_{\rm st}$ and $N_{\rm sc}$ around $\widetilde{\mu}=0$ in the spin-orbit-coupled cases shown in Fig.~\ref{fig:tc-effective-mu-components}(b2, c2). Moreover, the large DOS near the band bottom provides abundant low-energy fermionic states into which pairs can thermally dissociate. Heating can therefore produce a large increase in $N_{\rm F0}$, and hence in $R(T_{\rm c})$ defined in Eq.~(\ref{particle-production-rate}), when $\widetilde{\mu}$ is close to zero. Since the peak of $R(T_{\rm c})$ lies close to that of $C_V(T_{\rm c})$ in Fig.~\ref{fig:tc-cv-n0-response}, this enhanced thermal particle-number response provides a natural explanation for the specific-heat peak near $\widetilde{\mu}=0$.

In the case of the ordinary BCS--BEC crossover ($\lambda=0$), pairing fluctuations are enhanced by a strong pairing interaction: $N_{\rm sc}$ in Fig.~\ref{fig:tc-effective-mu-components}(a2) is larger than in the spin-orbit-coupled cases in (b2, c2) in the positive-$\widetilde{\mu}$ regime. In addition, since the DOS in a free Fermi gas with $\lambda=0$ behaves as $N(\omega)\propto\sqrt{\omega}$ (see the dash-dotted line in Fig.~\ref{fig:single-particle-dos}), a large value of $R(T_{\rm c})$ in Eq.~(\ref{particle-production-rate}) is expected not at $\widetilde{\mu}=0$ but at $\widetilde{\mu}>0$, where pairing fluctuations described by $N_{\rm sc}$ are enhanced to some extent. As a result, $C_V(T_{\rm c})$ is peaked at $\widetilde{\mu}>0$, where a stable molecule still does not exist.

Before ending this section, we briefly note that the DOS in the 3D spherical SOC case in Fig.~\ref{fig:single-particle-dos} is similar to the DOS in a 1D free Fermi gas with no SOC. While there is a threshold pairing-interaction strength below which a two-body bound state is not formed in a 3D Fermi gas, such a threshold is known to be absent in one and two spatial dimensions~\cite{Chadan_LowDimensionalBinding}. Thus, the 1D-like DOS in the present 3D spherical-SOC system is expected to promote the formation of two-body bound states even for a weak attraction, consistent with Ref.~\cite{Vyasanakere_TwoBodyBinding}. At finite density, this SOC-enhanced pair formation provides the two-body basis for the BCS--rashbon BEC crossover found in Ref.~\cite{Vyasanakere_Rashbons}. Indeed, while the BCS--rashbon BEC crossover also occurs in the system with a 2D SOC ($\lambda_x=\lambda_y=\lambda>0$, $\lambda_z=0$ in Eq.~(\ref{411a-plamb})), the low-energy DOS is constant, as in a 2D free Fermi gas~\cite{Cappelluti_PlanarSOCDOS} (see the dashed line in Fig.~\ref{fig:single-particle-dos}). Such a crossover phenomenon does not occur in the 1D SOC case ($\lambda_x=\lambda_y=0$, $\lambda_z>0$), where the DOS remains unchanged from that in a 3D free Fermi gas~\cite{Vyasanakere_TwoBodyBinding}.
\par
\par
\section{Specific heat in the BCS--rashbon BEC crossover region above $T_{\rm c}$}

In a previous paper~\cite{vanWyk_SpecificHeatTemperatureDependence}, one of the authors introduced two characteristic temperatures from the behavior of $C_V(T\geq T_{\rm c})$ in the BCS--BEC crossover regime of a Fermi gas, to draw the crossover phase diagram shown in Fig.~\ref{fig:normal-state-regime-cuts}(a). In this section, we extend this to the spin-orbit coupled case, to examine the normal-state properties above $T_{\rm c}$.

Before examining the BCS--rashbon BEC crossover case, we briefly review how to draw Fig.~\ref{fig:normal-state-regime-cuts}(a) from $C_V(T)$: Starting from the weak-coupling BCS regime, the low-temperature behavior of $C_V(T\geq T_{\rm c})$ gradually deviates from the specific heat $C_V^{\rm F}(T)$ in a free Fermi gas with increasing pairing-interaction strength, as shown in Figs.~\ref{fig:ordinary-cv-temperature}(a) and \ref{fig:ordinary-cv-temperature}(b).

\begin{figure}[tbp]
  \centering
  \paperfigure{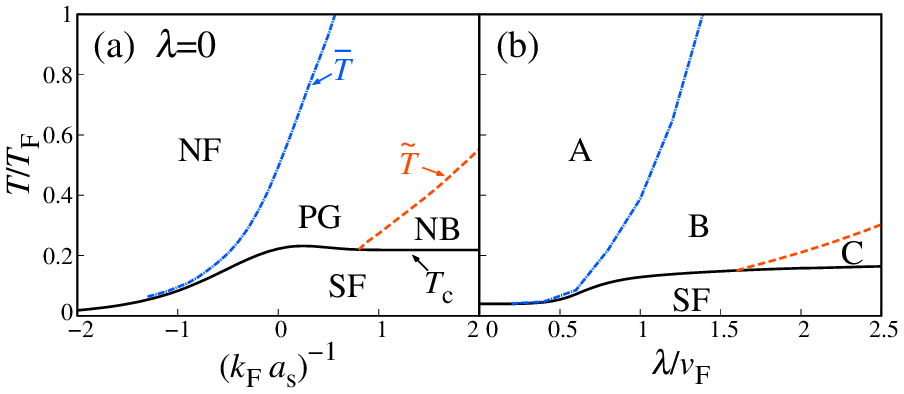}{Evaluated characteristic temperatures $\overline{T}$ and $\widetilde{T}$ in (a) the BCS--BEC crossover case ($\lambda=0$) and (b) the BCS--rashbon BEC crossover case, where $(k_{\rm F}a_{\rm s})^{-1}=-1.5$. In panel (a), NF, NB, and PG denote the normal Fermi gas regime, normal Bose gas regime, and pseudogap regime, respectively. SF denotes the superfluid phase in both panels. In panel (b), A, B, and C label the three normal-state regions separated by $\overline T$ and $\widetilde T$. In this figure, $T_{\rm c}$ is the only phase-transition temperature. $\overline{T}$ and $\widetilde{T}$ are crossover temperatures without being accompanied by any phase transition.}{fig:normal-state-regime-cuts}
\end{figure}

\begin{figure}[tbp]
  \centering
  \paperfigure{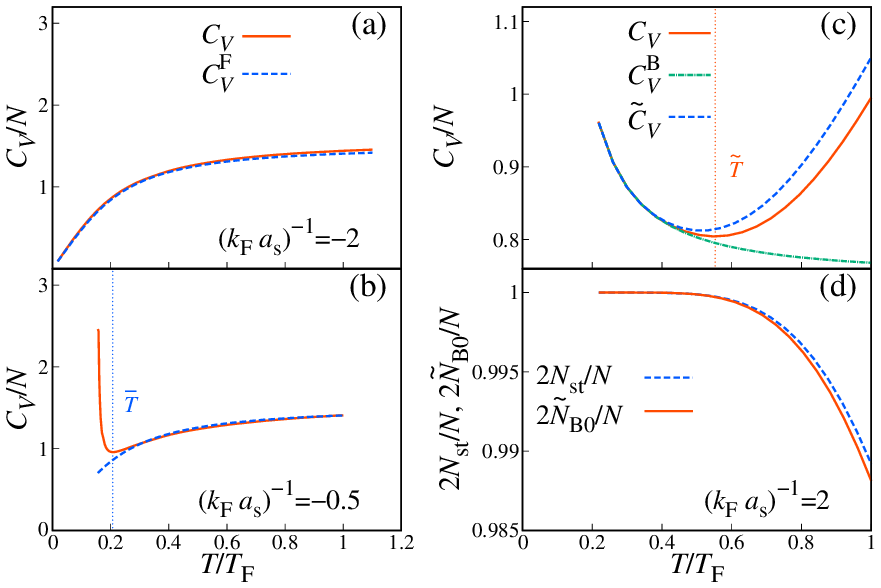}{(a)--(c) Calculated $C_V(T\geq T_{\rm c})$ in the BCS--BEC crossover regime ($\lambda=0$). Panel (d) compares the stable-pair number $N_{\rm st}$ in the NSR theory with the boson number $\widetilde N_{\rm B0}$ in the atom--molecule model for the parameters of panel (c). $\overline{T}$ ($\widetilde{T}$) is determined as the temperature at which $C_V(T)$ takes a minimum value in the BCS (BEC) regime. $C_V^{\rm F}$ and $C_V^{\rm B}$ are the specific heat in a free Fermi gas and an ideal gas of $N/2$ molecular bosons with a molecular mass $M_{\rm B}=2m$. $\widetilde{C}_V$ is the specific heat in a model Bose--Fermi mixture with $E_{\rm bind}^{\rm 2b}(>0)$, and $\widetilde{N}_{\rm B0}$ is the number of bosons in this model. Panels (a), (b), and (c,d) correspond to $(k_{\rm F}a_{\rm s})^{-1}=-2$, $-0.5$, and $2$, respectively.}{fig:ordinary-cv-temperature}
\end{figure}

The enhancement of $C_V(T)$ near $T_{\rm c}$ in Fig.~\ref{fig:ordinary-cv-temperature}(b) is consistent with the pairing-fluctuation interpretation discussed in Sec.~III. Following Ref.~\cite{vanWyk_SpecificHeatTemperatureDependence}, we use the temperature $\overline T$ at which $C_V(T)$ takes a minimum to distinguish the normal Fermi gas regime from the fluctuation-dominated regime. Although $\overline{T}$ is not the phase boundary, it is still convenient to distinguish between the normal Fermi gas regime (NF) and the so-called pseudogap regime (PG), where pairing fluctuations are important.

In the strong-coupling BEC regime, Fig.~\ref{fig:ordinary-cv-temperature}(c) shows that $C_V(T)$ near $T_{\rm c}$ agrees well with the specific heat $C_V^{\rm B}(T)$ in an ideal gas of $N/2$ Bose molecules with a molecular mass $M_{\rm B}=2m$. However, the difference between the two becomes appreciable when the number $N_{\rm st}$ of stable molecules deviates from $N/2$ (see Fig.~\ref{fig:ordinary-cv-temperature}(d)), indicating the onset of their thermal dissociation. Thus, to identify the region where the system may be regarded as a normal Bose gas (NB), we introduce the other characteristic temperature $\widetilde{T}$ at which $C_V(T)$ takes a minimum value in the BEC regime. Evaluating $\widetilde{T}$ and $\overline{T}$ at various pairing-interaction strengths, we obtain Fig.~\ref{fig:normal-state-regime-cuts}(a).

We note that the upturn behavior of $C_V(T)$ seen in the strong-coupling regime in Fig.~\ref{fig:ordinary-cv-temperature}(c) can also be obtained by considering a gas mixture of $\widetilde{N}_{\rm F0}$ free atoms and $\widetilde{N}_{\rm B0}$ Bose molecules having a molecular mass $M_{\rm B}=2m$, keeping the total fermion number $N=\widetilde{N}_{\rm F0}+2\widetilde{N}_{\rm B0}$ fixed, where

\begin{align}
  \widetilde{N}_{\rm F0}
  &=2\sum_{\bm p}n_{\rm F}\left(\widetilde{\xi}_{\bm p}^{\rm F}\right),\nonumber
  \\
  \widetilde{N}_{\rm B0}
  &=\sum_{\bm q}n_{\rm B}\left(\widetilde{\xi}_{\bm q}^{\rm B}\right),
  \label{eq:mixture-populations}
\end{align}
with
\begin{align}
  \widetilde{\xi}_{\bm p}^{\rm F}
  &=\frac{\bm p^2}{2m}-\mu=\varepsilon_{\bm p}^{\rm F}-\mu,\nonumber
  \\
  \widetilde{\xi}_{\bm q}^{\rm B}
  &=\frac{\bm q^2}{2M_{\rm B}}-E_{\rm bind}^{\rm 2b}-2\mu
   =\varepsilon_{\bm q}^{\rm B}-2\mu,
  \label{eq:mixture-dispersions}
\end{align}
and $E_{\rm bind}^{\rm 2b}=1/(ma_{\rm s}^2)$ being the binding energy of a two-body bound molecule~\cite{Chin_FeshbachResonance}. The internal energy $\widetilde{E}$ of this model is simply given by

\begin{align}
  \widetilde{E}
  =2\sum_{\bm p}\varepsilon_{\bm p}^{\rm F}
  n_{\rm F}\left(\widetilde{\xi}_{\bm p}^{\rm F}\right)
  +\sum_{\bm q}\varepsilon_{\bm q}^{\rm B}
  n_{\rm B}\left(\widetilde{\xi}_{\bm q}^{\rm B}\right).
  \label{eq:mixture-energy}
\end{align}

Then, as shown in Fig.~\ref{fig:ordinary-cv-temperature}(c), the calculated specific heat $\widetilde C_V(T)$ in this model qualitatively reproduces the upturn of $C_V(T)$ above $\widetilde T$, supporting the interpretation that thermal dissociation contributes to this behavior. In a sense, this model is a simplified version of the NSR theory in the BEC regime, where the scattering-state contribution $N_{\rm sc}$ is completely ignored. Thus, the difference between $C_V$ and $\widetilde{C}_V$ for $T\gtrsim\widetilde{T}$ may reflect the omitted scattering-state contribution.

\begin{figure}[tbp]
  \centering
  \paperfigure{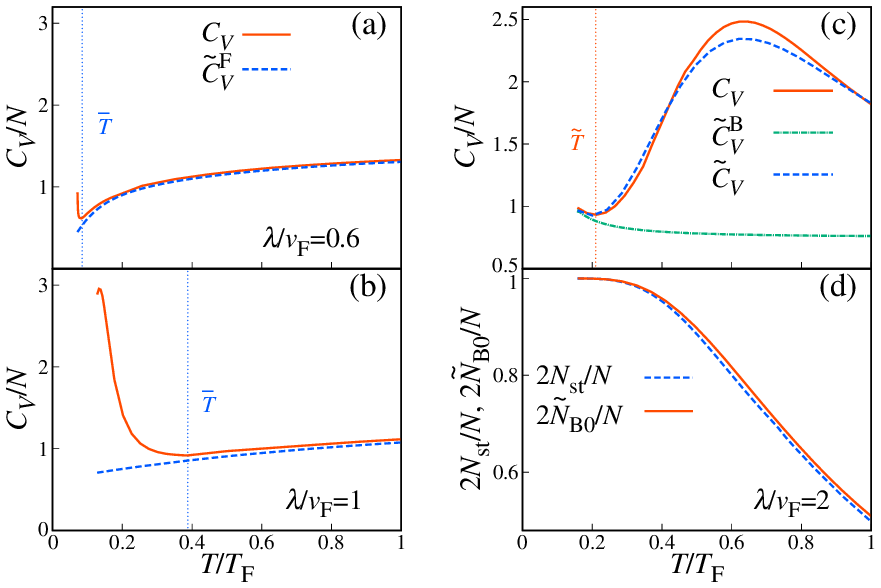}{Calculated $C_V(T)$ in the normal state above $T_{\rm c}$. Panels (a) and (b) show the cases in which $\overline T$ is identified, while panel (c) shows the case in which $\widetilde T$ is identified. Panel (d) compares the stable-rashbon number $N_{\rm st}$ obtained within the NSR theory with the boson number $\widetilde{N}_{\rm B0}$ in the atom--molecule model for the same parameters as in (c). $\widetilde{C}_V^{\rm F}$ is the specific heat in a non-interacting Fermi gas with the spherical SOC. $\widetilde{C}_V^{\rm B}$ is the specific heat of a gas of $N/2$ rashbons with a mass $M_{\rm RB}$ in Eq.~(\ref{app-mrb}). $\widetilde{C}_V$ and $\widetilde{N}_{\rm B0}$ are the result of the model atom--molecule mixture given in the text where a free Fermi gas and a molecular Bose gas are replaced by a spin-orbit-coupled non-interacting Fermi gas and a rashbon Bose gas, respectively. All panels have $(k_{\rm F}a_{\rm s})^{-1}=-1.5$.}{fig:soc-cv-temperature}
\end{figure}

We now proceed to the spin-orbit-coupled case shown in Fig.~\ref{fig:soc-cv-temperature}. In panels (a) and (b), pairing fluctuations enhance $C_V(T)$ near $T_{\rm c}$, and this enhancement is suppressed on heating, as in Fig.~\ref{fig:ordinary-cv-temperature}(b). In panel (c), $C_V(T)$ near $T_{\rm c}$ instead follows the ideal-rashbon-Bose-gas result, as in the molecular BEC case in Fig.~\ref{fig:ordinary-cv-temperature}(c). The subsequent increase accompanies the depletion of stable pairs shown in Fig.~\ref{fig:soc-cv-temperature}(d) and develops into a broad maximum. Above this maximum, the specific-heat enhancement is reduced as pair correlations are thermally depleted, consistently with the suppression of the enhancement in panels (a) and (b). As in the molecular BEC case discussed above, this behavior can be interpreted in terms of thermal dissociation: the atom--molecule model also reproduces the broad maximum and subsequent decrease. Using the same criteria as in the BCS--BEC crossover, we determine $\overline{T}$ and $\widetilde{T}$ in the spin-orbit-coupled case. Then, we obtain the crossover phase diagram at, for example, $(k_{\rm F}a_{\rm s})^{-1}=-1.5$, shown in Fig.~\ref{fig:normal-state-regime-cuts}(b).

Figure~\ref{fig:normal-state-regime-cuts}(b) shows that $\overline{T}$ and $\widetilde{T}$ divide the normal-state region into three, as in the BCS--BEC crossover case ($\lambda=0$) shown in Fig.~\ref{fig:normal-state-regime-cuts}(a). When we determine these characteristic temperatures in the whole $(k_{\rm F}a_{\rm s})^{-1}-(\lambda/v_{\rm F})$ plane to see how Figs.~\ref{fig:normal-state-regime-cuts}(a) and \ref{fig:normal-state-regime-cuts}(b) are related to each other, we obtain Fig.~\ref{ttilde-tbar}. We then find that “A,” “B,” and “C” in Fig.~\ref{fig:normal-state-regime-cuts}(b) correspond to the normal Fermi gas regime (NF), pseudogap regime (PG), and normal Bose gas regime (NB) known in the BCS--BEC crossover case.

In the BCS--BEC crossover case, it has been shown that the so-called pseudogap appears in DOS~\cite{Tsuchiya_HarmonicTrap,Perali_PseudogapLimit} due to strong fluctuations in the Cooper channel. Since $C_V$ is not so sensitive to this phenomenon, we need to directly evaluate single-particle excitations in the presence of SOC to clarify whether or not it also appears in the ``pseudogap'' regime in the BCS--rashbon BEC crossover case, which we leave for future work.

Regarding this, when we apply the above-mentioned atom--molecule model to the strong-SOC regime by replacing free Fermi atoms (molecules) with spin-orbit-coupled ones (rashbons), the model more closely approximates the NSR result above $\widetilde{T}$ (that is, in the ``pseudogap'' regime) than in the strong-coupling BEC case (see $\widetilde{C}_V$ and $\widetilde{N}_{\rm B0}$ in the right panels in Figs.~\ref{fig:ordinary-cv-temperature} and \ref{fig:soc-cv-temperature}). The origin of this difference remains to be clarified. One possible explanation is that, for the weak attraction considered in the BCS--rashbon BEC crossover, correlations between unpaired Fermi atoms are weaker than in the strong-coupling BEC case. These correlations contribute to the scattering-state term $N_{\rm sc}$ in the NSR number equation (\ref{number-eq}) and are omitted in the atom--molecule model. If their contribution is smaller, the normal state above $\widetilde T$ may be more closely approximated by a mixture of Fermi atoms and rashbons.

\begin{figure}[tbp]
  \centering
  \paperfigure{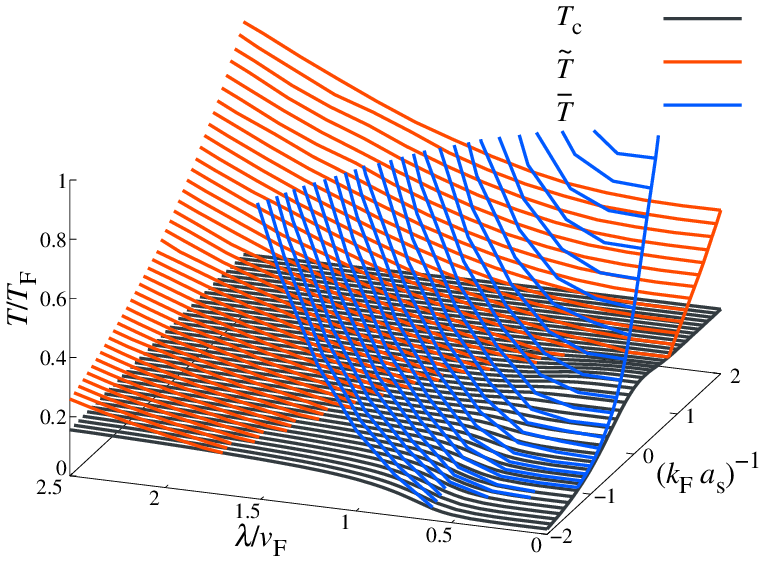}{Superfluid transition temperature $T_{\rm c}$ and characteristic temperatures $\widetilde T$ and $\overline T$ as functions of $(k_{\rm F}a_{\rm s})^{-1}$ and $\lambda/v_{\rm F}$.}{ttilde-tbar}
\end{figure}
\par
\section{Summary}
\par
To summarize, we have examined normal-state properties of an interacting Fermi gas with a 3D spherical SOC. Including effects of SOC on the single particle dispersion, as well as pairing fluctuations within the strong-coupling theory developed by Nozi\`eres and Schmitt-Rink, we calculated the specific heat at constant volume $C_V(T=T_{\rm c})$ as a function of the pairing-interaction strength and SOC strength.
\par
Starting from the weak-coupling BCS regime, $C_V$ at $T_{\rm c}$ was shown to exhibit a peak structure with increasing SOC strength, to eventually approach the same value as the specific heat in an ideal rashbon Bose gas at $\widetilde{T}_{\rm BEC}$ in the strong-SOC limit. This BCS--rashbon BEC crossover behavior of $C_V(T_{\rm c})$ is similar to the pairing-interaction dependence of this quantity in the BCS--BEC crossover regime, where $C_V(T_{\rm c})$ is also peaked in the unitary regime. However, while the strong pairing interaction is essentially important in the latter case, effects of SOC on single-particle excitations, giving a 1D-like DOS, play a crucial role in the former.
\par
In the BCS--BEC crossover regime of an ultracold Fermi gas, the normal state above $T_{\rm c}$ can be divided into the normal Fermi gas regime, pseudogap regime, and normal Bose gas regime on the basis of detailed analyses of the temperature dependence of $C_V(T)$. In this paper, we extended this approach to the spin-orbit coupled system, to draw the crossover phase diagram in terms of the pairing interaction strength, SOC strength, and the temperature.
\par
Although we have clarified how these three regions expand in the presence of 3D spherical SOC, it remains an interesting future problem to clarify how SOC affects the physical properties of each region. For example, in the BCS--BEC crossover phenomenon, it is known that the strong pairing interaction gives rise to the so-called pseudogap in DOS~\cite{Tsuchiya_HarmonicTrap,Perali_PseudogapLimit} and spin-gap in the susceptibility~\cite{Tajima_SpinGap}. At this stage, however, it is unclear whether or not such phenomena also occur in the corresponding regime in the BCS--rashbon BEC crossover when the pairing interaction itself is weak.
\par
In addition, since the BCS--rashbon BEC crossover also occurs in a 2D SOC, it also remains to be studied how the 2D SOC alters the results obtained in this paper. Furthermore, we only dealt with the normal state at and above $T_{\rm c}$ in this paper. Extending this work to the superfluid phase below $T_{\rm c}$ would therefore also be important. The BCS--BEC crossover has been studied in $^{40}\mathrm{K}$ and $^{6}\mathrm{Li}$ Fermi gases by tuning the pairing interaction through a Feshbach resonance. Exploring an alternative route to this crossover is therefore an important issue in cold Fermi gas physics. Regarding this, the BCS--rashbon BEC crossover realized by tuning a synthetic SOC is a promising idea, and our results could be useful for the future development of this research field.
\par
\par
\begin{acknowledgments}
We thank H. Tsukada, H. Takeda, and T. Suzuki for useful discussions. This work was supported by JST SPRING, Japan Grant Number JPMJSP2123. Y. Ohashi was supported by a Grant-in-Aid for Scientific Research from MEXT and JSPS (Grants No. 22K03486 and No. 26K06983).
\end{acknowledgments}
\par
\appendix

\begingroup
\section{Decomposition of $N_{\rm NSR}$ into $2N_{\rm st}$ and $N_{\rm sc}$}
\label{app:nst-nsc}
The spectral representation of the NSR correction term $N_{\rm NSR}$ in Eq.~(\ref{number-nsr}) is given in Ref.~\cite{SaDeMelo_FiniteTemperatureFluctuations} as
\begin{align}
  N_{\rm NSR}
  =
  \sum_{\bm q}
  \int_{-\infty}^{\infty}d\omega\,
  n_{\rm B}(\omega)
  A(\bm q,\omega),
  \label{app-nnsr-spectral}
\end{align}
where
\begin{align}
  A(\bm q,\omega)
  =
  -\frac{1}{\pi}\operatorname{Im}
  \left[
    \Gamma(\bm q,\omega+i\delta)
    \pdif{}{\mu}
    \Pi(\bm q,\omega+i\delta)
  \right].
  \label{app-pair-spectral-density}
\end{align}
$A(\bm q,\omega)$ has the meaning of the bosonic single-particle spectral weight.

When a real pole of $\Gamma(\bm q,i\nu_n\to\omega+i\delta)$ appears below the threshold energy of the two-particle continuum given by
\begin{align}
  \omega_{\rm th}(\bm q)
  =
  \min_{\bm p,s,s'}
  \left[
    \xi^s_{\bm p+\frac{\bm q}{2}}
    +
    \xi^{s'}_{\bm p-\frac{\bm q}{2}}
  \right],
  \label{app-continuum-threshold}
\end{align}
it describes a stable (rashbon) bound state. In this case, $\Gamma(\bm q,i\nu_n\to\omega+i\delta)$ has the form near the pole at $\omega=\omega_{\bm q}$
\begin{align}
  \Gamma(\bm q,\omega+i\delta)
  \simeq
  \frac{1}
  {
    (\omega+i\delta-\omega_{\bm q})
    \left.
      \pdif{}{\omega}
      \Pi(\bm q,\omega)
    \right|_{\omega=\omega_{\bm q}}
  }.
  \label{app-pole-expansion}
\end{align}
One then finds that Eq.~(\ref{app-pair-spectral-density}) near this pole has the form
\begin{align}
  A^{\rm pole}(\bm q,\omega)
  =
  \delta(\omega-\omega_{\bm q})
  \frac{
    \left.
      \pdif{}{\mu}
      \Pi(\bm q,\omega)
    \right|_{\omega=\omega_{\bm q}}
  }{
    \left.
      \pdif{}{\omega}
      \Pi(\bm q,\omega)
    \right|_{\omega=\omega_{\bm q}}
  }.
  \label{app-pole-spectral-density}
\end{align}
Here, $\delta(x)$ is the Dirac delta function.
Summing up the contribution of such stable poles to Eq.~(\ref{app-nnsr-spectral}) gives the number $N_{\rm st}$ of stable molecules (rashbons for $\lambda>0$) in Eq.~(\ref{nst-pole-number}), that is
\begin{align}
  N_{\rm st}
  =
  \frac{1}{2}
  \sum_{\bm q:\,\mathrm{pole}}
  n_{\rm B}(\omega_{\bm q})
  \frac{
    \left.
      \pdif{}{\mu}
      \Pi(\bm q,\omega)
    \right|_{\omega=\omega_{\bm q}}
  }{
    \left.
      \pdif{}{\omega}
      \Pi(\bm q,\omega)
    \right|_{\omega=\omega_{\bm q}}
  }.
  \label{app-nst-pole}
\end{align}
Here, the $\bm q$ summation is taken over stable poles $\omega_{\bm q}$.

The scattering-state contribution $N_{\rm sc}$ in Eq.~(\ref{nst-nsc-decomp}) is then given as the remaining part of $N_{\rm NSR}$:
\begin{align}
  N_{\rm sc}
  =
  N_{\rm NSR}-2N_{\rm st}.
  \label{app-nsc-continuum}
\end{align}
\endgroup

\par
\section{Derivation of Eqs.~(\ref{energy_of_rb}) and (\ref{number_of_rb})}
\label{app:strong-coupling}

In the strong-SOC bound-pair limit, the Fermi chemical potential $\mu$ lies below the bottom of the lower-helicity band, $-m\lambda^2/2$~\cite{He_SphericalSOCMass,Yamaguchi_SphericalSOCNSR}. We consider temperatures sufficiently low that $\min_{\bm p}\xi^-_{\bm p}=-\mu-m\lambda^2/2\gg T$. Under this condition, the Fermi distribution functions in the pair correlation function $\Pi(\bm q,i\nu_n)$ in Eq.~(\ref{pair-correlation}) are exponentially suppressed and can be neglected. In this case, the pole equation of the analytically continued particle-particle scattering matrix $\Gamma(\bm q,i\nu_n\to\omega+i\delta)$,
\begin{align}
  0
  ={}&
  \frac{m}{4\pi a_{\rm s}}
  -
  \frac{1}{4}
  \sum_{\bm p,s,s'=\pm}
  \frac{
    1+ss'
    \dfrac{
      \left(\bm p+\frac{\bm q}{2}\right)_\lambda\cdot
      \left(\bm p-\frac{\bm q}{2}\right)_\lambda
    }{
      \left|\left(\bm p+\frac{\bm q}{2}\right)_\lambda\right|
      \left|\left(\bm p-\frac{\bm q}{2}\right)_\lambda\right|
    }
  }{
    \omega+i\delta
    -\xi^s_{\bm p+\frac{\bm q}{2}}
    -\xi^{s'}_{\bm p-\frac{\bm q}{2}}
  }
  -\sum_{\bm p}\frac{m}{p^2}
  \label{app-gamma-real-strong}
\end{align}
reduces to the pole equation for a rashbon bound state in Ref.~\cite{He_SphericalSOCMass}, when $\omega$ is regarded as the rashbon dispersion $\omega_{\bm q}$ measured from its chemical potential.

In the case of spherical SOC, $\omega_{\bm q}$ is given in Ref.~\cite{He_SphericalSOCMass} up to $O(q^4)$ as
\begin{align}
  \omega_{\bm q}
  =
  \frac{\bm q^2}{2M_{\rm RB}}
  -\mu_{\rm RB}
  +O(q^4).
  \label{app-pole-dispersion}
\end{align}
Here,
\begin{align}
  \mu_{\rm RB}
  =
  2\mu+E_{\rm bind}^{\rm RB},
  \label{app-murb}
\end{align}
is the rashbon chemical potential, and
\begin{align}
  M_{\rm RB}
  =
  \frac{2m}{
    \dfrac{7}{3}
    -\dfrac{2}{J}
    -\dfrac{4}{3}
    \left[
      \dfrac{J-1}{J}
    \right]^{3/2}
  }
  \label{app-mrb}
\end{align}
is the rashbon mass, where
\begin{align}
  J
  =
  1+\frac{1}{4}
  \left[
    \frac{1}{m\lambda a_{\rm s}}
    +
    \sqrt{
      \left[\frac{1}{m\lambda a_{\rm s}}\right]^2+4
    }
  \right]^2.
  \label{app-j-definition}
\end{align}
In Eq.~(\ref{app-murb}),
\begin{align}
  E_{\rm bind}^{\rm RB}
  =
  m\lambda^2J
  \label{app-ebind}
\end{align}
is the binding energy of a rashbon.
Expanding the denominator of $\Gamma(\bm q,\omega+i\delta)$ around the rashbon bound state, one has
\begin{align}
  \Gamma(\bm q,\omega+i\delta)
  \simeq
  \frac{Z_{\rm RB}}{
    \omega+i\delta
    -\dfrac{\bm q^2}{2M_{\rm RB}}
    +\mu_{\rm RB}
  },
  \label{app-gamma-real}
\end{align}
where $Z_{\rm RB}$ is the residue of the rashbon pole. Returning to the Matsubara frequency, we obtain
\begin{align}
  \Gamma(\bm q,i\nu_n)
  \simeq
  \frac{Z_{\rm RB}}{
    i\nu_n
    -\dfrac{\bm q^2}{2M_{\rm RB}}
    +\mu_{\rm RB}
  }.
  \label{app-gamma-bose}
\end{align}
Thus, in the bound-pair limit, the scattering matrix has the same low-energy form as the Green's function of a boson with mass $M_{\rm RB}$ and chemical potential $\mu_{\rm RB}$. The deviation of $M_{\rm RB}$ from $2m$ reflects the SOC-modified internal and center-of-mass motion of the bound pair~\cite{He_SphericalSOCMass,Vyasanakere_CollectiveExcitations}. Substitution of this pole form into $\Omega_{\rm NSR}$ yields the ideal Bose-gas thermodynamic potential. This is the strong-coupling composite-boson limit of BCS--BEC crossover theories~\cite{Ohashi_ResonantNSRTheory,Haussmann_SelfConsistentLimit,Pieri_CompositeBosonLimit}.

Substituting Eq.~(\ref{app-gamma-bose}) into $\Omega_{\rm NSR}$ in Eq.~(\ref{omega-nsr}), and dropping unimportant $T$-independent terms, one reaches
\begin{align}
  \Omega_{\rm NSR}
  &\simeq
  T\sum_{\bm q,i\nu_n}e^{i\delta\nu_n}
  \ln\!\left(
    i\nu_n
    -\frac{\bm q^2}{2M_{\rm RB}}
    +\mu_{\rm RB}
  \right)
  \nonumber\\
  &=
  T\sum_{\bm q}
  \ln\!\left[
    1-\exp\!\left[
      -\beta\left(
        \frac{\bm q^2}{2M_{\rm RB}}-\mu_{\rm RB}
      \right)
    \right]
  \right].
  \label{app-omega-rb}
\end{align}

Under the same condition $-\mu-m\lambda^2/2\gg T$, the non-interacting contribution $N_{\rm F0}$ is thermally suppressed, and the number equation~(\ref{number-eq}) reduces to
\begin{align}
  N
  \simeq
  N_{\rm NSR}
  =
  -\left(
    \pdif{\Omega_{\rm NSR}}{\mu}
  \right)_T
  \simeq
  2\sum_{\bm q}
  n_{\rm B}\!\left(
    \frac{\bm q^2}{2M_{\rm RB}}-\mu_{\rm RB}
  \right),
  \label{app-number-rb}
\end{align}
where Eq.~(\ref{app-murb}) was used in carrying out the $\mu$ derivative.

The internal energy is also obtained in the same manner. Ignoring the non-interacting part $\Omega_0$, we obtain
\begin{align}
  E
  &\simeq
  \Omega_{\rm NSR}
  -T\left(
    \pdif{\Omega_{\rm NSR}}{T}
  \right)_\mu
  -\mu\left(
    \pdif{\Omega_{\rm NSR}}{\mu}
  \right)_T
  \nonumber\\
  &=
  \sum_{\bm q}
  \frac{\bm q^2}{2M_{\rm RB}}
  n_{\rm B}\!\left(
    \frac{\bm q^2}{2M_{\rm RB}}-\mu_{\rm RB}
  \right)
  -E_{\rm bind}^{\rm RB}\frac{N}{2}.
  \label{app-energy-rb}
\end{align}

\end{document}